\documentclass[12pt,letterpaper]{article}
\usepackage{comment}
\usepackage[sort&compress]{natbib}
\usepackage{graphicx}
\usepackage{wrapfig}
\usepackage{bm}
\renewcommand{\refname}{\hfil References Cited\hfil}                   

\def\bR{{\bf R}}
\def\br{{\bf r}}
\def\bk{{\bf k}}

\def\bv{{\bf v}}
\def\beq{\begin{equation}}
\def\eeq{\end{equation}}
\def\bea{\begin{eqnarray}}
\def\eea{\end{eqnarray}}
\usepackage{amssymb}
\usepackage{amsmath}
\usepackage{epsfig}
\usepackage{subfigure}

\begin{document}
\setcounter{page}{1}

\noindent
{\centerline{{\bf BCS-BEC Crossover and the Unitary Fermi Gas}}}
\bigskip

\noindent
{\centerline{Mohit Randeria$^1$ and Edward Taylor$^2$}}

\medskip\noindent
{\centerline{$^1$The Ohio State University}}
\noindent
{\centerline{$^2$McMaster University}}
\bigskip
\bigskip

\noindent
{\bf Abstract:} The crossover from weak coupling Bardeen-Cooper-Schrieffer (BCS) pairing 
to a Bose-Einstein condensate (BEC) of tightly bound pairs, as a function of the attractive interaction 
in Fermi systems, has long been of interest to theoretical physicists. The past decade has seen a 
series of remarkable experimental developments in ultracold Fermi gases that has realized the
BCS-BEC crossover in the laboratory, bringing with it fresh new insights into the very strongly
interacting unitary regime in the middle of this crossover. In this review, we start with a pedagogical introduction
to the crossover and then focus on recent progress in the strongly interacting regime. While our focus
is on new theoretical developments, we also describe three key experiments that
probe the thermodynamics, transport and spectroscopy of the unitary Fermi gas.
We discuss connections between the unitary regime and other areas of physics
-- quark-gluon plasmas, gauge-gravity duality and high temperature superconductivity --
and conclude with open questions about strongly interacting Fermi gases.
\bigskip
\bigskip


\noindent
{\underline{{\bf 1. Introduction:}}}
\medskip

The problem of the BCS-to-BEC crossover first arose in an attempt to understand 
superconductivity and superfluidity going beyond the standard paradigms.
Recent developments in trapping, cooling and controlling the interactions in ultracold Fermi gases have led to a realization
of the BCS-BEC crossover in the laboratory. The most interesting new developments, both
in theory and experiment, relate to a very strongly interacting state of matter -- the unitary Fermi gas -- that 
lies right at the heart of this crossover. 

The goal of this review is to provide an introduction to the BCS-BEC crossover and describe some of the beautiful new results on 
the unitary Fermi gas. Our main focus will be on theoretical developments, although we will also describe some experimental results
as the interplay between theory and experiment has been central to recent developments.

Until recently, all known superfluids and superconductors
fell into one of two disjoint classes: bosonic and fermionic. This, in fact, led to two quite distinct paradigms , BEC and BCS, 
for understanding the ``super'' properties of quantum fluids.

The BEC paradigm, first developed for non-interacting bosons and later generalized to
take into account repulsive interactions, describes bosonic fluids like $^4$He or ultracold Bose gases like $^{87}$Rb.
The condensate is a macroscopic occupation of a single quantum state that occurs below a transition temperature $T_c$,
which, even in strongly interacting Bose systems like $^4$He, is of the same order of magnitude as
the quantum degeneracy temperature at which the inter-particle spacing becomes of the order
of the thermal de Broglie wavelength. Although the bosons studied in the laboratory are composite
particles made up of an even number of fermionic constituents, this internal structure is quite
irrelevant for the low energy properties of the superfluid ($T < T_c$) or normal ($T > T_c$) states.

The BCS paradigm, first developed for metallic superconductors, describes a pairing instability arising
from a weak attractive interaction in a highly degenerate system of fermions. The formation of pairs
and their condensation both occur at the same $T_c$ that is orders of magnitude smaller than the Fermi energy
$E_F$, which sets the scale of the degeneracy temperature (we set $\hbar = k_B = 1$).
The BCS theory is not only spectacularly successful in describing conventional superconductors, and predicting new phenomena
(like the Josephson effects), it has also been generalized to describe a variety of systems including pairing in nuclei and fermionic superfluidity in $^3$He.

Several recent developments, both in experiment and theory, have led to an examination of fermionic superconductors and superfluids
that cannot be adequately described within the BCS framework. Perhaps the most notorious of these is
the still unsolved problem of high temperature superconductivity in the cuprates, which we will briefly return to in the Concluding section.
Our main focus, however, is on the BCS-BEC crossover. Early theoretical work on the crossover
was of conceptual interest, but the real excitement came from 
its experimental realization in 2004--2005~\cite{regal04SF,zwie04rescond,kina04sfluid,bour04coll,chin04gap,part05,zwie05vort}.
There is now a clear recognition that the BCS and BEC paradigms are not as distinct as they were once thought to be, but rather are the two extrema of a 
continuum. It is particularly interesting that right in the middle of this crossover lies a most strongly interacting state
-- the unitary Fermi gas -- which has several remarkable properties. The unitary regime has the highest ratio of $T_c/E_F \simeq 0.15 - 0.2$
ever observed in any fermionic superfluid, a large pairing energy gap $\Delta \simeq 0.5 E_F$, and an unusual
normal state with an anomalously low shear viscosity to entropy density ratio $\eta/s \sim 0.2$ that comes close to saturating
a lower bound derived in a very different context using gauge-gravity duality in string theory.

For a fuller account of the subject of this review and a more complete set of references, we recommend Ref.~\cite{zwerger2011bcs}, which contains over a dozen chapters
written by leading experts.  Other review articles include Refs.~\cite{kett08varenna, bloc08review, gior08review,levin05review}.
In the remainder of this Section, we give a brief overview of the prehistory of theoretical studies of the BCS-BEC crossover,
prior to the cold atoms era, and of the key experimental developments in the early days of ultracold Fermi gas experiments.
We conclude the Section with an outline of the rest of the article.

\medskip
\noindent
{\bf History:} 
The idea of invoking some sort of BEC to understand superconductivity is an old one dating back to 
Schafroth \textit{et al.}~\cite{scha57}.  However, their theory had problems in dealing with real metals, not least because the
Cooper pairs are hugely overlapping in real space -- hardly describable as point bosons -- and the antisymmetry of the electronic wavefunction 
was crucial to the problem of superconductivity. The BCS description in terms of ``momentum-space pairing'', on the other hand, faced up to all these challenges
and made quantitative predictions for the properties of superconductors known at the time, in addition to giving deep new insights.
In the wake of the success of BCS theory, the differences with BEC were stressed much more often than their commonalities.

In a pioneering paper, Eagles~\cite{eagl69} studied superconductivity in doped semiconductors like SrTiO$_3$
with a very low carrier density, where the attraction between electrons need not be 
small compared with the Fermi energy $E_F$. This led to the first mean-field treatment of the
BCS-BEC crossover. (The problem of superconductivity in doped SrTiO$_3$ is back in vogue again
with recent experiments on oxide interfaces.) 

Independently, Leggett~\cite{legg80} addressed the problem of the BCS-BEC crossover in a dilute gas of fermions at $T=0$
motivated by superfluid $^3$He. Although $^3$He is very much in the BCS limit, Leggett wanted to
understand the extent to which some of its properties, such as the total angular momentum of the
superfluid, might be similar to that of a BEC of diatomic molecules.
A finite temperature analysis of the BCS-BEC crossover, along with the
evolution of the critical temperature $T_c$ was
first presented by Nozi\`eres and Schmitt-Rink (NSR)~\cite{nozi85}.
We will come back to the results of Leggett and NSR in Section 3.

With the discovery of high temperature superconductors in 1986 and the
realization that the pair size is only slightly larger than the average interparticle spacing, there was a resurgence of interest in
the BCS-BEC crossover. A simple model like a Fermi gas with  a strong attractive
$s$-wave interaction is of course unable to quantitatively describe the high $T_c$ materials, where 
$d$-wave superconductivity arises upon doping an antiferromagnetic Mott insulator.
Nevertheless, important new ideas like that of a pairing pseudogap that were introduced in these
investigations are relevant for all short coherence-length superconductors. These ideas will be discussed below.

Questions analogous to the BCS-BEC crossover also arise in
several other problems of condensed matter physics. These include the
problem of exciton condensation, as first discussed by Keldysh~\cite{keld95}, although long range Coulomb interactions make 
the physics somewhat different. Another problem of great interest is 
that of the crossover from a weak-coupling Slater insulator with spin density wave antiferromagnetism to the 
strong-coupling Mott insulator with local moment antiferromagnetism in the repulsive Hubbard model at half-filling. 
In fact the mathematical description of this crossover from itinerant to local moment magnetism is quite similar to the BCS-BEC crossover 
in a lattice problem such as the attractive Hubbard model.  In this review we will focus exclusively on 
continuum, as opposed to lattice, formulations of the crossover since all of the experimental activity to date has
been on Fermi gases without an underlying lattice. It is likely, however, that as progress is made on cooling
fermions in optical lattices, we will soon see experiments on the weak-to-strong coupling crossovers in both
repulsive and attractive Fermi Hubbard models.  Finally, we note that very similar ideas have been explored in the context of colour superconductivity~\cite{alfo00color}.  

\medskip
\noindent
{\bf Ultracold Fermi gas experiments:} 
Ultracold Fermi gases began to be studied experimentally soon after the 1995 observation of BEC in ultracold
Bose gases. Cooling Fermi gases turned out to be harder than their bosonic counterparts, but quantum degeneracy in an
atomic Fermi gas was first established in 1999~\cite{dema99}. It was soon realized that one could exploit 
the Feshbach resonance (see Section 2) to tune the effective interaction between two hyperfine species of fermions, the two states
being the analogs of spin $|$$\uparrow\rangle$ and  $|$$\downarrow\rangle$ electrons in a superconductor. 
Unlike in the bosonic case where increasing the strength of interactions
leads to loss of stability due to enhanced three-body losses, in the Fermi gas, Pauli exclusion saves the day~\cite{petr04dimers}.
This leads to previously unexpected stability in the strongly interacting regime of Fermi gases, which has the highest superfluid 
transition temperatures and the most interesting properties, but which was never before studied in the laboratory.

\begin{figure}   
\centering
\includegraphics[width=0.6\textwidth]{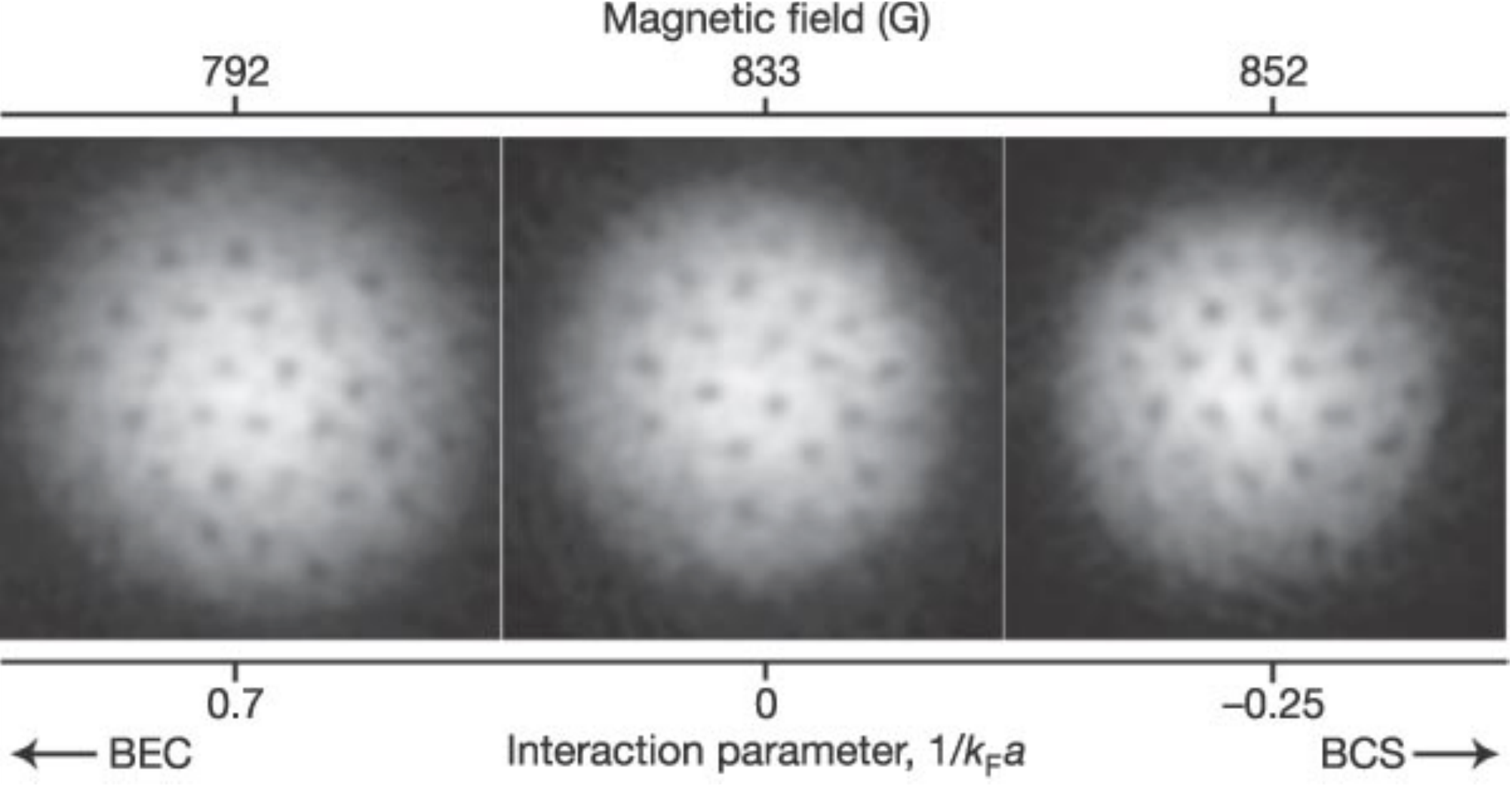}
\caption{Quantized vortices in rotating Fermi gases.  From Ref.~\cite{zwie05vort}.} 
\label{vortex-fig}
\end{figure}

Observing the transition to the superfluid phase for fermions also turned out to be harder than in the Bose gases, since 
the phase transition does not lead, in general, to a significant change in the density profile of a trapped gas~\cite{regal04SF,zwie04rescond}. The first experiments
demonstrating a superfluid transition were in the BEC regime where the fermions are already bound into bosonic diatomic molecules above $T_c$.
The condensation of these molecular bosons has a signature similar to that of atomic BEC.
An ingenious pair-projection technique was introduced to probe condensation in the fermionic regime, wherein the 
fermionic BCS pairs were projected onto molecular pairs and then imaged~\cite{regal04SF}. Once condensation of fermion pairs was
demonstrated at $T_c/E_F \simeq 0.15 - 0.2$, the experimental (and theoretical) activity in this field exploded.
We conclude this subsection by mentioning only one out of many beautiful experiments: a direct proof of
superfluidity was provided by the observation of an Abrikosov vortex lattice in a rotating Fermi gas; see Fig.~\ref{vortex-fig}.

\medskip
\noindent
{\bf Outline:} 
The rest of this review is organized as follows.  In Section 2, we introduce the Feshbach resonance and 
the $s$-wave scattering length $a$, the tunable interaction in Fermi gases.
We then turn in Section 3 to the simplest description -- mean-field theory plus Gaussian fluctuations -- that provides a 
qualitatively correct picture of the entire crossover.  
We next describe recent theoretical progress in the strongly interacting, unitary
regime where $a$ diverges, and the approximations of Section 3 are the least reliable. 
In Section 4, we describe field theoretical approaches and quantum Monte Carlo results
and in Section 5, we focus on exact results that are valid across the entire crossover, including
unitarity. In Section 6, we turn to three important experiments 
that probe the thermodynamics, transport and spectroscopy of the unitary Fermi gas.
These illustrate important aspects of the strongly interacting regime: the superfluid phase
transition, the anomalously low viscosity and the possibility of a pairing pseudogap.
In Section 7, we discuss some problems related to the BCS-BEC crossover (that we do not have space for in the main text) as well as some open questions.  We end with some concluding thoughts in Section 8. 

\bigskip
\noindent
{\underline{{\bf 2. Tunable interactions:}}}
\medskip

The Fermi atoms used in the BCS-BEC experiments have so far been either $^6$Li or $^{40}$K. Typical experimental parameters are:
total number of atoms $N \sim10^5 - 10^7$, inter-particle distance or $k_F^{-1}$ on the order of a micron, Fermi energy $E_F$ of order
$100$ nanoKelvin, temperatures going down to $\sim 0.1 E_F$. The two species of fermions are actually
two different hyperfine states, but are often called ``spin'' $|$$\uparrow\rangle$ and  $|$$\downarrow\rangle$ in accordance with the standard
usage of BCS theory. For experimental details, the reader is referred to Ref.~\cite{kett08varenna}.

The most important difference with all previously studied superfluids is the fact that the interaction between
$|$$\uparrow\rangle$ and  $|$$\downarrow\rangle$ fermions can be tuned in the laboratory. 
The average separation between atoms $k_F^{-1}$ is much larger than the range of the inter-atomic interaction potential $r_0$. 
For a {\it dilute gas}, with $k_F r_0 \ll 1$, in three dimensions the interaction can be specified by a single parameter, the {\it $s$-wave scattering length} $a$.
All thermodynamic and transport properties of  such a dilute gas can be written in a ``universal'' scaling form; for instance the free energy $F$
at any temperature can be written in the form 
\begin{equation}
\label{eq:scaling}
F = N E_F \ {\cal F}\!\left({T/E_F},{1/k_F a}\right),
\end{equation} 
where ${\cal F}$ is a dimensionless scaling function. This result is ``universal'' in the
sense that it is independent of all microscopic details, provided $k_F r_0 \rightarrow 0$.  For instance, the pressure of $^6$Li at a given value of the interaction parameter $1/k_Fa$ and temperature is the same as in $^{40}$K, modulo negligible corrections set by $k_Fr_0$.  
Deviations from universality, e.g., differences between results for $^6$Li and $^{40}$K, would be expected when effects on the scale of the range $r_0$ become important.

\begin{figure}   
\centering
\mbox{\subfigure{\includegraphics[width=0.4\textwidth]{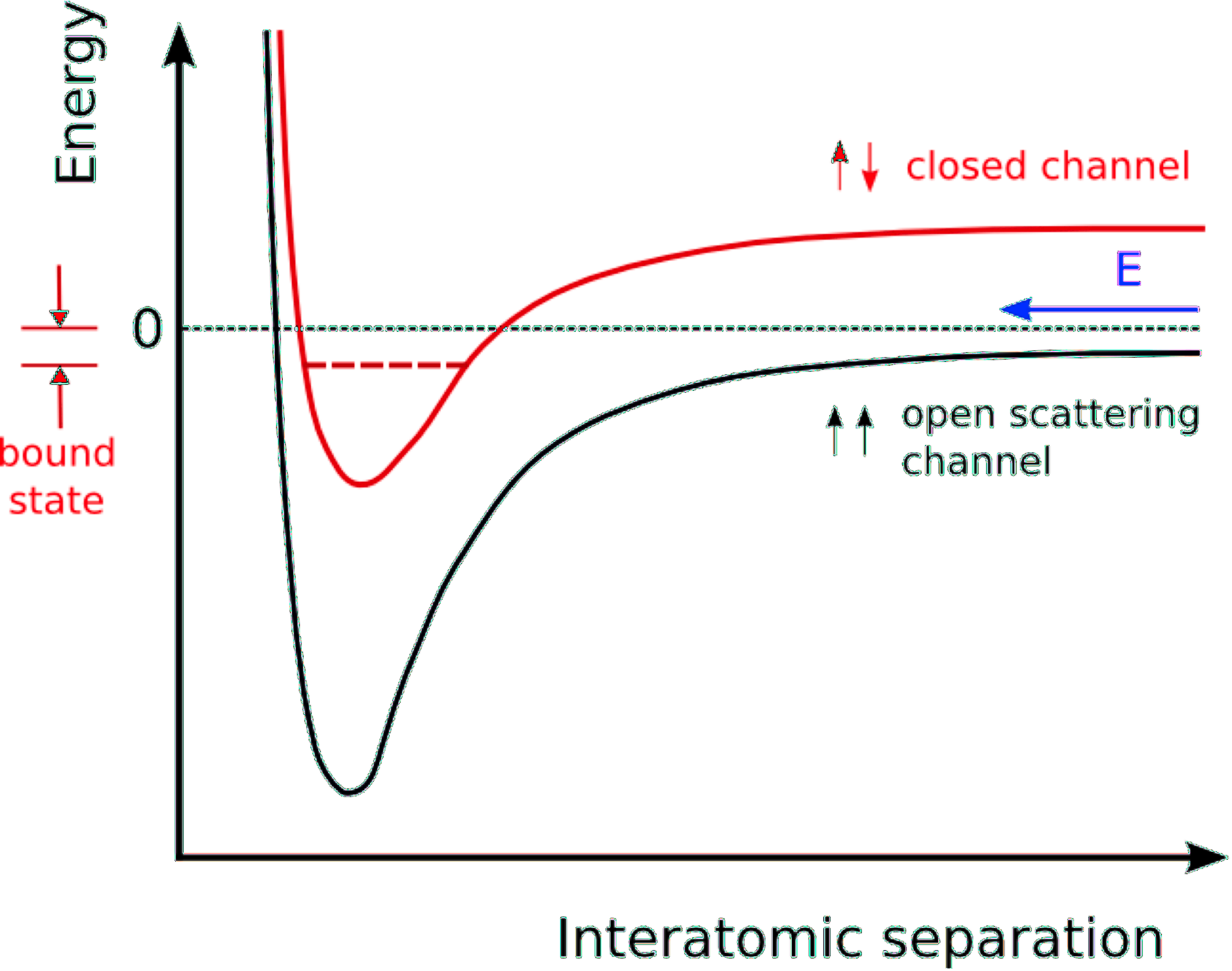}}\quad
\subfigure{\includegraphics[width=0.4\textwidth]{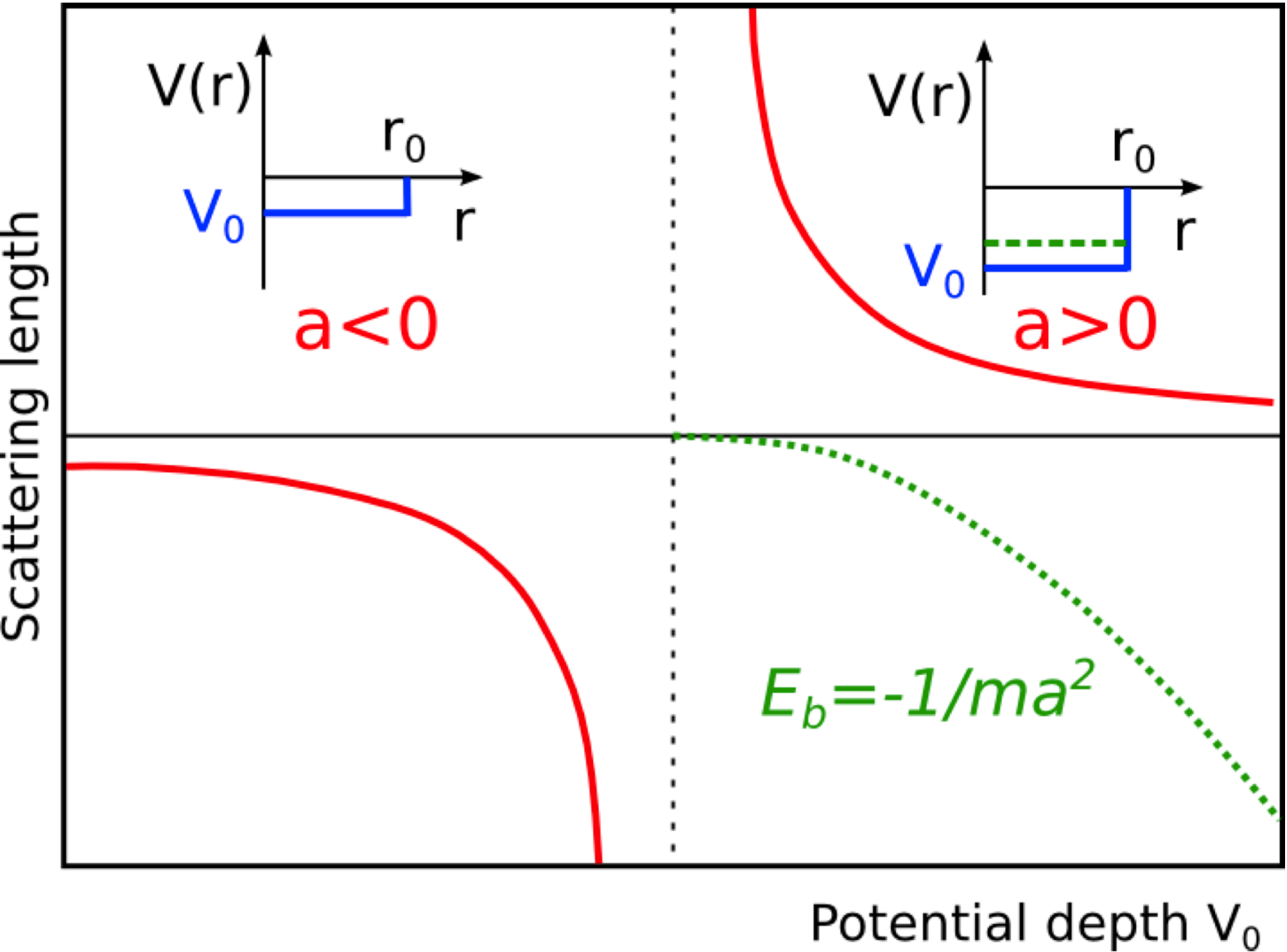}}}
\caption{Left panel: Feshbach resonance. Right panel: Scattering length for a square well potential 
and the appearance of a bound state of energy $E_b$ past a  critical well depth, signifying $|a|=\infty$.} \label{feshbachfig}
\end{figure}

Let us briefly describe the Feshbach resonance without going into too many technical details.
Consider the two-body problem in vacuum at $T\!=\!0$.
A Feshbach resonance is a dramatic increase in the collision cross-section
of two atoms when a bound state in the ``closed channel'' crosses
the scattering continuum of the ``open channel''; see the left panel of Fig.~\ref{feshbachfig}.
In the specific example of $^6$Li (electronic spin $S=1/2$, nuclear spin $I=1$) 
the electron spin is essentially fully polarized (at the magnetic fields of interest $B \geq 500\rm\,G$) and aligned in the same direction for
each of the three lowest hyperfine states. Thus two colliding $^6$Li atoms are in a continuum, spin triplet state
in the open channel. The closed channel has a singlet bound state state
that can resonantly mix with the open channel due to the hyperfine interaction that couples the
electron spin to the nuclear spin. (This is the only place in this article where we discuss the electronic spin of the Fermi
atoms; at all other places ``spin-up'' and ``-down'' denote the two hyperfine states of the two-species Fermi gas).

The difference in the magnetic moments in the closed and open channels allows the
experimentalist to use an external magnetic field $B$ as a knob to tune across a Feshbach resonance;
for details see Refs.~\cite{duin04feshbachreview,chin10feshbachreview,kett08varenna}. 
The resulting interaction between atoms in the open channel can be described by 
a $B$-dependent scattering length which, in the vicinity of a resonance, has the form
\begin{equation}
\label{eq:scattering-length}
a(B)=a_{\rm BG}\left[1-{\Delta B}/({B-B_0})\right].
\end{equation}
Here $a_{\rm BG}$ is the background value
in the absence of the coupling to the closed channel,
while $B_0$  and $|\Delta B|$ are the location and width of the resonance.
For most resonances of interest to experimentalists, $a_{\mathrm{BG}}\Delta B >0$, with the result that for increasing $B$ the inverse scattering length goes from positive to negative.  Thus
experimental results are often plotted as a function of  $-1/k_F a$ (increasing $B$) rather than $1/k_Fa$.

To get an intuitive feel for the scattering length, we do not need to understand the intricacies of the two-channel
model of a Feshbach resonance. Instead, we can look at the much simpler single-channel
problem of two particles with a short-range interaction.
This simplified discussion is quite sufficient to understand much of 
the current experimental and theoretical literature on cold Fermi gases. The technical reason for the validity of this
single-channel model is that most of the experiments are in the
so-called ``broad'' resonance limit where the effective range (which we do not discuss here) of the Feshbach resonance 
is much smaller than $k_F^{-1}$~\cite{bruu04univ,diener05broad,romans05}.  This ensures that the fraction of closed-channel molecules is extremely small, a feature directly confirmed in 
experiments~\cite{part05}.  

Consider then the problem of two fermions with ``spin'' $|$$\uparrow\rangle$ and  $|$$\downarrow\rangle$ 
interacting with a two-body potential with range $r_0$.
The low energy properties at momentum $k$, such that $k r_0 \ll 1$, are described by the $s$-wave 
scattering amplitude
\begin{equation}
\label{eq:scattering-amplitude}
f(k) = {1 \over{k\cot\delta_0(k)-ik}} \approx {-1 \over{1/a + ik}}.
\end{equation}
Here $\delta_0(k\to 0) = -\tan^{-1}(ka)$ is the $s$-wave scattering phase shift whose low-energy
behavior is completely determined by the scattering length $a$.

Since this effective interaction is independent of the detailed shape of the potential,
we can examine it for the simplest model potential --
a square well of depth $V_0$ and range $r_0$ -- to get a better feel for the scattering length $a$ 
as a function of $V_0$.
As shown in the right-panel of Fig.~\ref{feshbachfig}, $a < 0$ for weak attraction, 
grows in magnitude with increasing $V_0$, and diverges to $-\infty$
at the threshold for the formation of a two-body bound state in vacuum.
(The threshold for a square well is $V_0 = \pi^2/mr_0^2$, where the reduced mass is $m/2$.)
Once this bound state is formed, the scattering length changes sign and decreases
from $+\infty$ with increasing $V_0$. Above threshold, $a > 0$ has the simple physical 
interpretation as the size of the bound state, whose energy is given by $-1/ma^2$.

The threshold for bound state formation in the two-body problem, where 
$|a|\rightarrow \infty$, is called the \emph{unitary} point. Here the phase shift
$\delta_0(k=0) = \pi/2$ and the scattering amplitude $f \approx -1/ik$ takes its
maximum value allowed by unitarity. As we shall see, the 
many-body problem at unitarity is the most strongly interacting 
regime in the BCS-BEC crossover.

In addressing the many-body problem, it is more convenient to use a ``zero-range" contact potential in the Hamiltonian
\beq 
{\cal{H}} = \psi^{\dagger}_{\sigma}\left(-{\nabla^2}/{2m}-\mu\right)\psi_{\sigma} - g(\Lambda)\psi^{\dagger}_{\uparrow}\psi^{\dagger}_{\downarrow}\psi_{\downarrow}\psi_{\uparrow}
\label{H}
\eeq
to incorporate the relevant $s$-wave scattering physics.  Here the chemical potential $\mu$ controls the density of fermions with dispersion $\epsilon_{\bk} = k^2/2m$.
The attraction between the two ``spin'' species is characterized by a coupling $g(\Lambda)$, where the inverse range of the potential
determines the cutoff $\Lambda \simeq 1/r_0 \gg k_F^{-1}$. We choose the ``bare'' $g(\Lambda)$ such that it leads to a 
``renormalized'' interaction described by the scattering length $a$; see e.g., Refs.~\cite{rand90,melo93}. This is given by the two-particle Schr\"odinger equation in vacuum, 
written in ${\bf k}$-space as 
\beq 
\frac{m}{4\pi a} = \frac{-1}{g(\Lambda)} + \sum_{|\bk|<\Lambda}\frac{1}{2\epsilon_{\bk}}.
\label{LS}
\eeq
The ultraviolet divergence associated with the zero-range potential in the many-body problem is then
regularized using Eq.~(\ref{LS}), with $\Lambda\!\rightarrow\!\infty$ in the calculation of any observable.  

An equivalent real-space approach is often very useful.
The $s$-wave wavefunction for the relative motion of two particles in vacuum, whose energy
vanishes at infinity, has the form
\beq 
\psi(r) \propto \left(1/r-1/a\right)\;\;\mathrm{for}\;\; r \geq r_0
\label{2particle}\eeq
The $N$-particle wavefunction in the many-body problem must then have the same short-distance
behavior when any two particles with opposite ``spins'' come together, keeping the remaining $(N-2)$ particles fixed. 

\medskip
\noindent
{\underline{{\bf 3. Global phase diagram:}}}
\medskip

We are now ready to address the many-body problem of a finite density
of ``spin'' $\uparrow$ and  $\downarrow$ fermions 
with a two-body interaction specified by the scattering length $a$, so
that the dimensionless coupling constant is $1/k_F a$.
The BCS limit $1/k_F a \rightarrow -\infty$ corresponds to a weak attraction that
is not able to form a two-body bound state in vacuum, but nevertheless leads to a 
collective Cooper instability in the presence of a Fermi surface.
In the opposite limit $1/k_F a \rightarrow +\infty$, strong attraction leads to
tightly bound diatomic molecules which exhibit BEC. 
In this Section we focus primarily on the simplest theoretical approaches that qualitatively
describe the evolution from BCS to BEC. In addition to giving physical insight, this discussion will help the 
reader better appreciate recent progress, both quantitative and conceptual, that is described later in
the review.

\begin{figure}   
\begin{center}
\includegraphics[width=0.6\textwidth]{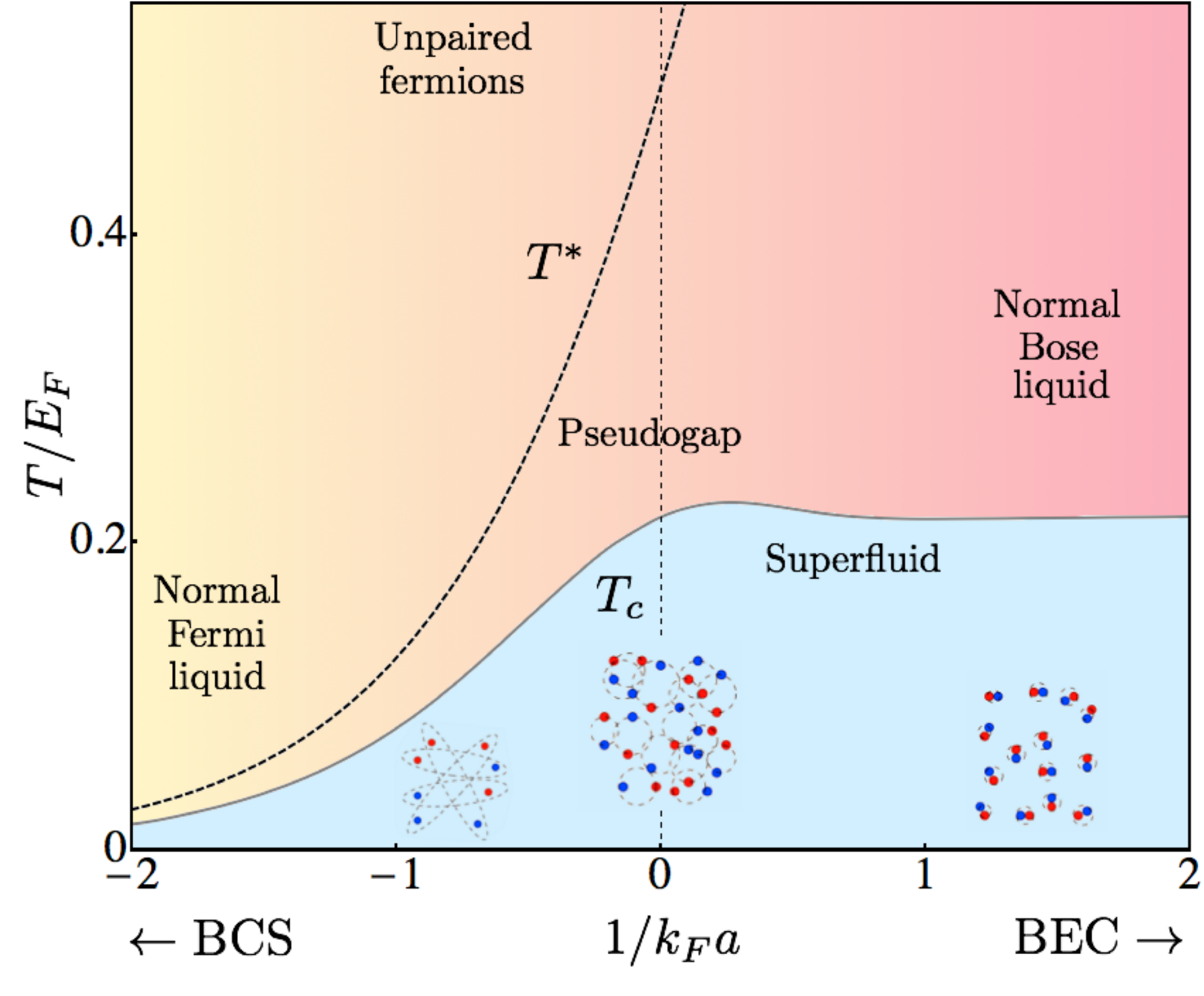}
\caption{
Qualitative phase diagram of the BCS-BEC crossover as a function of temperature $T/E_F$
and coupling $1/k_F a$, where $k_F$ is the Fermi momentum and $a$ the scattering length (based on the results of~\cite{melo93}).  
The pictures show schematically the evolution from the BCS limit with large Cooper pairs to the
BEC limit with tightly bound molecules.  Unitarity ($1/k_Fa = 0$) corresponds to strongly interacting pairs with size comparable to $k_F^{-1}$.
The pair-formation crossover scale $T^*$ diverges away from the transition temperature
$T_c$ below which a condensate exists and the system is superfluid, as the attraction increases.  The best quantitative estimates of $T_c$ and $T^*$ along with the question of the pseudogap at unitarity are discussed in the text.  
}\label{f:phasediagram}
\end{center}
\end{figure}

\medskip
\noindent
{\bf Ground state crossover:} It was recognized early on~\cite{eagl69,legg80} 
that the BCS wavefunction continues to provide a qualitatively reasonable variational description of the pairing correlations 
for arbitrary attraction. The $T=0$ crossover mean field theory (MFT) is essentially the
same as standard BCS theory, except that one also has to self-consistently determine
the renormalization of the chemical potential $\mu$ together with the 
MFT gap equation. $\mu$ decreases monotonically with increasing attraction, going from $E_F$ 
in the BCS limit to a {\it negative} value in the BEC limit, where it  approaches half the pair binding energy, $-1/2ma^2$.

MFT describes a smooth crossover from the weak-coupling BCS limit with large, overlapping Cooper pairs 
of size $\xi_{\rm pair} \sim k_F^{-1}\exp(+\pi/2 k_F |a|) \gg k_F^{-1}$ all the way to
the strong-coupling BEC regime of tightly bound dimers ($\xi_{\rm pair} \ll k_F^{-1}$).
There is no singularity in the many-body state at unitarity
(the threshold for a two-body bound state in vacuum) because collective
Cooper pairs have already formed at arbitrarily weak attraction in the many-body problem.
Mathematically, the divergence of $a$ at unitarity does not lead to singular behavior
since all observable are functions of $1/k_Fa$ as already noted in Eq.~(\ref{eq:scaling}).

It is important to emphasize that although the attraction increases monotonically from
BCS to BEC, both limits are weakly interacting. This is obvious for the BCS limit. It is also
true for the BEC limit because once the strong attraction is resolved by dimer formation,
all that remains is a weak residual {\it repulsion} between dimers,  which vanishes in the deep BEC limit;
see below.

The most strongly interacting regime is right in the middle of the BCS-BEC crossover where the
scattering length $|a| \gg k_F^{-1}$. In the ground state at unitarity $1/k_F a= 0$, the pair size 
is on the order of the interparticle spacing $\xi_{\rm pair} \simeq k_F^{-1}$. We will see in later Sections
that some of the most exciting new developments have been in the unitary regime.

\medskip
\noindent
{\bf Finite temperature properties:} To determine the $T>0$ phase diagram, one
needs to determine the superfluid transition temperature $T_c$ as a function of $1/k_Fa$. 
MFT~\cite{melo93} yields the BCS result $T_c = (8 \gamma/\pi e^2)E_F \exp(-\pi/2 k_F |a|)$
when $1/k_F a \rightarrow -\infty$. 
With increasing attraction, the MFT $T_c$ estimate becomes qualitatively incorrect. 
In fact, as argued in Ref.~\cite{melo93}, the MFT estimate is really a ``pairing temperature'' $T^*$ below
which a significant fraction of fermions are bound in pairs and which has
nothing to do with condensation except in the weak coupling BCS regime.
In the strong coupling limit $T^* \sim |E_b|/\ln(|E_b|/E_F)$~\cite{melo93} is the Saha dissociation temperature
for dimers with binding energy $E_b = -1/ma^2$.

Nozi\`eres and Schmitt-Rink (NSR)~\cite{nozi85} gave a diagrammatic approximation 
for calculating $T_c$ that interpolates smoothly between the exponentially
small BCS result and the non-interacting BEC $T_c = [n/2\zeta(3/2)]^{2/3}\pi/m \simeq 0.22 E_F$
for dimers of mass $2m$ and density $n/2$; see Fig.~\ref{f:phasediagram}.  Just as MFT is a
saddle point approximation in a functional integral formulation of the crossover,
NSR is equivalent to Gaussian fluctuations in the
normal state~\cite{melo93,engel97becbcs,rand95}.

Any perturbative treatment of fluctuations will necessarily break down in the (Ginzburg) critical region near $T=T_c$.  
While the range of $|T-T_c|/T_c$ where critical fluctuations dominate is small in both the BCS
and BEC limits, it is of order unity near unitarity~\cite{engel97becbcs,taylor09crit}.  
This follows from the non-monotonic dependence on $1/k_Fa$ of the Ginzburg--Landau (GL) healing length, which is large in both the BEC and BCS limits and has a minimum $\sim k^{-1}_F$ near unitarity.   The GL healing length (also called the coherence length)  $\xi_{\mathrm{GL}} \equiv \sqrt{\gamma/|\alpha|}$ is defined in terms of the parameters entering the GL free energy functional $f_{\mathrm{GL}}= \alpha |\Psi|^2 + \gamma|\nabla \Psi|^2 + \cdots$, where $\Psi$ is the superfluid order parameter. 
We should emphasize that $\xi_{\mathrm{GL}}$ is in general distinct~\cite{engel97becbcs} from the Cooper pair size, which is a monotonically decreasing function of $1/k_Fa$.  Related to the non-monotonicity of $\xi_{\mathrm{GL}}$ through the crossover is the non-monotonic dependence on $1/k_F a$ of the 
critical velocity $v_c$, which is determined by pair-breaking in the BCS regime but by the speed of sound in the BEC regime.
It has a maximum at unitarity with $v_c\!\sim\!v_F$~\cite{sens06vort,comb06coll}, a prediction that has been 
directly verified by experiment~\cite{mill07critical}.  

Although the $T=0$ superfluid is maximally robust close to unitarity, $T_c$ as a function of $1/k_F a$ shows
only a rather weak maximum and then flattens out in the BEC regime, as seen in Fig.~\ref{f:phasediagram} and also in
QMC simulations discussed below. In the BCS regime $T_c$ is determined by the gap, while in the BEC regime it is controlled by the
superfluid density, or phase stiffness, whose scale is set by the density in our Galilean invariant system.
$T_c$ for a lattice model of the crossover, such as the ``negative U'' Hubbard model,  however, has a strong maximum~\cite{paiva10} and
decreases like $t^2/|U|$ in the BEC regime.
  
The normal (non-superfluid) state crossover is more subtle than 
the ground state crossover from large to small pairs. 
In the BCS limit, since both pair formation and condensation occur at $T_c\ll E_F$, 
the normal state is a Landau Fermi liquid.
In the BEC regime, on the other hand, superfluid order is destroyed
by phase fluctuations depleting the condensate, {\it not} by destroying pairing. The state above $T_c$  
is a normal Bose gas of dimers, which dissociate only at the pairing $T^*$.
The question of how the system above $T_c$ evolves from a normal Fermi liquid to a normal
Bose liquid is quite nontrivial. It was proposed early on that it does so via a
pairing pseudogap~\cite{rand92,triv95,rand98varenna} 
between $T_c$ and $T^*$.
The existence of a pseudogap would be particularly exciting near unitarity where the system can be
in a degenerate Fermi regime and yet show marked deviations from Fermi-liquid behavior; see Sec.~7.

\medskip
\noindent
{\bf Beyond the simplest approaches:}
While the results described above have the virtue of interpolating smoothly between the BCS and BEC limits,
there is no small parameter to control the calculations in the strongly interacting regime. The results, though
qualitatively correct, are quantitatively inadequate.  
Even in the BCS limit, MFT fails in two distinct ways. First, MFT overestimates both the $T\!=\!0$ gap $\Delta$
and $T_c$ by the same numerical factor of $(4e)^{1/3}\simeq 2.2$. (The $T_c$ prefactor  is reduced 
from $8 \gamma/\pi e^2 \simeq 0.61$ to $\simeq 0.28$).
This suppression arises from polarization effects in the medium (particle-hole fluctuations)
that effectively weaken the attraction~\cite{gork61}.
Second, the weak-coupling MFT ground state energy ignores  
perturbative corrections in $k_F|a|$, while including the exponentially small
non-perturbative contribution of pairing. The perturbative ``Fermi liquid'' corrections at $T\!=\!0$
are in fact correctly described by Gaussian fluctuations \cite{dien08} about MFT, and
have the same form as the classic results for the repulsive Fermi gas, but with $a < 0$.

In the opposite BEC limit $1/k_F a \gg 1$ the crossover 
MFT overestimates the dimer scattering scattering length $a_{dd}^{\rm MF}=2 a$~\cite{melo93},
while the exact solution of the four-body problem~\cite{petr04dimers} 
yields $a_{dd}=0.6a$.
Gaussian fluctuation theories \cite{hu06becbcs,dien08} that work across the entire crossover are able to 
partially account for this renormalization in the BEC limit.

It is no surprise that the quantitative failures of MFT and simple extensions are the most severe in the
strongly interacting regime near unitarity; see Table I. Consider $\xi_s = E_0/(3 N E_F/5)$ 
(subscript $s$ for superfluid), the ratio of the ground state energy $E_0$ at unitarity to that of the
free Fermi gas~\cite{bert00}. At unitarity, one also has $\mu = \xi_s E_F$. The MFT estimate $\xi_s^{MFT} = 0.59$~\cite{engel97becbcs} is much larger
than the best estimates $\xi_s \simeq 0.37$. 
Similarly, the MFT plus Gaussian fluctuation approach, which provides the estimate $T_c/E_F \simeq 0.2$~\cite{melo93} at unitarity,  
exceeds the best numerical estimates of $0.15$. Below, we show the best estimates for various quantities in Table I
and discuss where they come from.

Despite the lack of a small parameter, there have been 
many approximate calculations using diagrams~\cite{pera04bcsbec,levin05review,hu07universal} (see also, Ch.~4 in Ref.~\cite{zwerger2011bcs}),
functional integrals ~\cite{engel97becbcs,hu06becbcs,dien08}, and conserving approximations~\cite{haus94,haus07bcsbec}.
While the level of self-consistency imposed 
and the results differ in detail, the essential physics that these approximations attempt to capture is the same, namely
the effect of pair (particle-particle channel) fluctuations going beyond MFT. 
These calculations do give important qualitative insights, even if not precise quantitative control.   For instance, one can see the smooth evolution of the Anderson-Bogoliubov phonon mode in the weak coupling BCS limit to the Bogoliubov phonon in the BEC limit.  
The reduction in the ground state energy, 
discussed above in terms of $\xi_s$, can be attributed~\cite{dien08} to the zero point motion of these collective modes and virtual scattering of gapped quasiparticle excitations that are missing in MFT. The conserving approximation
results using Luttinger-Ward functionals~\cite{haus94,haus07bcsbec} are in quantitative agreement with the
most accurate results shown in Table I, except for a spurious first-order phase transition.

One way to organize these calculations is to generalize Eq.~(\ref{H}) to $N$ ``flavors'' of spin $\uparrow$ and  $\downarrow$ fermions
interacting with an $Sp(2N)$-invariant attractive interaction and carry out a large-$N$ expansion~\cite{veil07,niko07renorm}.
The $N=\infty$ saddle point corresponds to MFT and $1/N$ corrections to Gaussian fluctuations. In the end, however,
one has to take $N = 1$.
 
\medskip 
\noindent
{\underline{{\bf 4. Unitary Fermi gas: Field theories and Quantum Monte Carlo studies}}}
\medskip

The three-dimensional unitary Fermi gas is the most strongly interacting system of fermions with a short range 
interaction that one can possibly have in the continuum.
When $|a| = \infty$, the low-energy $s$-wave scattering phase shift $\delta_0 = \pi/2$ is the largest it can be, and
the scattering amplitude $f(k)  = i/k$ has no scale. 

The theoretical challenge then is to gain insight into properties of the unitary Fermi gas --
for instance, the ground state energy, $T_c$, and scaling functions -- in the absence of a small parameter. 
In recent years, there has been remarkable theoretical progress on addressing this problem, using
a variety of approaches. These include:
(1) diagrammatic and functional integral approaches (described above),
(2) field theory techniques -- renormalization group, $1/N$ and $\epsilon$ expansions, and operator product expansion -- which were developed for scale-invariant problems of classical and quantum critical phenomena,
(3) numerical simulations using various quantum Monte Carlo (QMC) methodologies, and 
(4) exact results for zero-range interactions that give nontrivial relations between
various physical observables. We describe (2) and (3) here and devote the next Section to the new exact results.

\medskip
\noindent
{\bf Field theory approaches:} Universality, arising from the lack of a scale other than $E_F$  at unitarity, was recognized
in Refs.~\cite{bert00,ho04uni}. This observation was formulated in a renormalization group (RG) framework in Ref.~\cite{niko07renorm}, 
which emphasized that the 3D unitary point at $T=0$ and $\mu=0$ is a quantum critical point.
At $T=0$, it separates a vacuum  phase with no particles for $\mu < 0$ from a superfluid phase for $\mu > 0$.
A suitably defined dimensionless coupling constant $g$, which determines the strength of the 
four fermion interaction $\psi^\dag_\uparrow\psi^\dag_\downarrow\psi_\downarrow\psi_\uparrow$, satisfies the RG equation
$dg/d\ell = (2-d)g - g^2/2$ where $d$ is the spatial dimensionality. In $d>2$ there is a fixed point at $g^* = 2(2-d)< 0$ that represents the scale invariant point. Couplings $g > g^*$ (less attractive than the fixed point) flow to $0$,
so that no bound state exists in vacuum, while  $g < g^*$ (more attractive than the fixed point) flow to $-\infty$, indicating the existence
of a bound state. $T$ and $\mu$ are also relevant perturbations at this fixed point. Standard RG arguments can then be used to 
write down various observables as functions of $(g^*-g)\!\sim\!1/a$ (in 3D), $T$ and $\mu$ in a scaling form like Eq.~(\ref{eq:scaling}).
Moreover, the RG also suggests strategies to compute scaling functions. One approach is to
look at a suitable large-$N$ limit of the problem, leading to the $1/N$-expansion discussed above.

Another powerful approach uses dimensionality expansions~\cite{nish06epsilon, nish07eps}.
It is easy to see from the above RG flow that
2D is the ``lower critical dimensionality'' with the fixed point at $g^* = 0$; i.e., the weak-coupling BCS limit. 
This is related to the fact that an arbitrarily weak attraction 
in 2D leads to a bound state, the implications of which for the BCS-BEC crossover were examined in Refs.~\cite{rand89bound,rand90}.
Thus, in $d= 2+\epsilon$ dimensions, the fixed point $g^*$ is perturbatively accessible and one can compute various observables
in powers of $\epsilon$. The ``upper critical dimension'' $d=4$~\cite{nuss06becbcs} is not immediately apparent from the formulation discussed here because the scale-invariant fixed point lies in the BEC limit where the relevant weakly-interacting degrees of freedom are bosonic.
Using a Hubbard-Stratonovich transformation, one can write the interaction in terms of 
a pair of fermions coupling to a bosonic field $\phi$ via $\lambda(\phi\psi^\dag_\uparrow\psi^\dag_\downarrow + {\rm h.c.})$,
similar to the two-channel formulation. This coupling $\lambda$ becomes dimensionless in $d=4$~\cite{nish06epsilon, nish07eps,niko07renorm} .
Another way to see that $d=4$ is special is to look at the pair wavefunction (\ref{2particle}), which has the asymptotic form  
$\psi(r\to 0)\sim 1/r^{d-2}$ in $d$-dimensions. One then sees that its normalization integral diverges for $d\geq 4$.  The amplitude of the pair wavefunction is clustered around the origin $r=0$ and the system looks like an ideal gas of point bosons in $d=4$~\cite{nuss06becbcs,nish07eps}.  
Starting from the upper critical dimension, the properties of the unitary Fermi gas are thus again perturbatively accessible in a $d = 4 - \epsilon$ expansion.
The situation is analogous to classical statistical mechanics, where also two distinct
formulations are needed near the lower and upper critical dimensionalities: a non-linear sigma model, with fixed-length spins,
near 2D and a $\varphi^4$ theory, with amplitude fluctuations, near 4D.

We refer the reader to the review article~\cite{NSinzwerger2011bcs}, which summarizes the results for thermodynamic 
and spectral properties using dimensionality expansions and their extrapolation to 3D (shown in Table I).
An alternative approach to describe the properties of the unitary Fermi gas using 
nonrelativistic conformal field theory~\cite{son06symmetry, wern06unitary, nish07CFT} is also described there.

\medskip
\noindent
{\bf Quantum Monte Carlo (QMC) results:} Numerical simulations using a variety of QMC techniques have played a decisive role in
providing quantitative insights into the unitary Fermi gas, as might be expected for a strongly interacting problem without
a small parameter. The earliest QMC papers focused on the evolution of the ground state 
properties~\cite{carl03,astr04} across the crossover using $T\!=\!0$ ``diffusion'' QMC. 
In this wavefunction-based technique one starts with a trial state, like the BCS ground state
with a fixed number of particles, possibly together with additional correlation factors. One then uses a QMC technique to project out,
as it were, the component of the true ground state and compute observables. 
The ``fermion sign problem" manifests itself here as the bias introduced by the fixed-node approximation,
since projection does not move the nodes of the trial states, at least in the simplest implementations. Nevertheless, this technique
has led to some of the most accurate estimates of the ground state energy. Early QMC studies at unitarity found
$\xi_s\!\simeq\!0.42\!-\!0.44$~\cite{carl03,astr04}, but more recent QMC gives a bound $\xi_s\!<\!0.383(1)$~\cite{forb10},
consistent with~\cite{bulg08thermo}, which uses a different QMC method.

Other QMC studies~\cite{carl08gap} find a Bogoliubov quasiparticle dispersion of the form  $E({\bf k}) =$ $\left[(\hbar^2k^2/2m^* -\tilde{\mu})^2+\Delta^2\right]^{1/2}$ at $T\!=\!0$,
where the energy gap $\Delta/E_F \simeq 0.5$ at unitarity~\cite{carl05,bulg08thermo}.
Here the effective mass $m^*$ and the shift in the minimum of $E({\bf k})$
arising from $\tilde{\mu} = \mu - U$ are due to self-energy effects.
The predicted negative shift $U \simeq - 0.43E_F$~\cite{carl08gap} and
the energy gap have been seen in radio frequency spectroscopy experiments~\cite{schi08gap}.

Determinental QMC techniques are used to address finite temperature properties.
In general, fermion QMC is plagued by the ``sign problem'': the determinant obtained upon
integrating out the fermions is not positive definite resulting in large cancellations
that lead to large statistical errors. Fortunately, it has long been known that for the special case
of spin-balanced fermions with on-site attractive interaction on a lattice, there exists a
Hubbard-Stratonovich decoupling that evades the fermion sign problem.
Using variants of the determinental QMC approaches, different groups have obtained
$T_c/T_F=0.152(7)$~\cite{buro06TC,buro08TC}, $0.15(1)$~\cite{bulg08thermo} and $0.171(5)$~\cite{goul10}
at unitarity. QMC studies also show a non-monotonic $T_c$ as a function of $1/k_F a$ with a peak on
the BEC side of unitarity~\cite{buro08TC}.  A comparison of  theoretical results with experiments at unitarity are shown in Table I.

\begin{center}
{\renewcommand{\arraystretch}{1.3}
    \begin{tabular}{| c | c | c | c |}
    \hline
    $|a|=\infty$& $\xi_s=E_0/({3\over 5}NE_F)$ & $\Delta/E_F$ & $T_c/E_F$ \\ \hline
   MFT ($T\!=\!0$)/NSR ($T_c$)  & 0.59~\cite{engel97becbcs} & 0.68~\cite{engel97becbcs} & 0.2~\cite{melo93} \\ \hline
   $\epsilon$-expansion~\cite{NSinzwerger2011bcs}& 0.377(14) & 0.60 & 0.180(12) \\ \hline
    QMC & $<$ 0.383(1)~\cite{forb10} & 0.5~\cite{carl05,bulg08thermo} & 0.152(7)~\cite{buro06TC,buro08TC} \\ \hline
     Experiment & 0.376(5)~\cite{ku12} & 0.44~\cite{schi08gap} & 0.167(13)~\cite{ku12} \\ \hline
         \end{tabular}}
        
        \bigskip
         {\bf Table I}: Ground state energy $E_0$, the $T\!=\!0$ gap $\Delta$, and  $T_c$ at unitarity. Error bars are given in
         $(\dots)$ and references in $[\ldots]$  Very recent high-precision measurements of the Feshbach resonance in $^6$Li suggest that $\xi_s$ might be slightly reduced~\cite{zurn13}.
\end{center}

\bigskip
\noindent
{\underline{{\bf 5. Exact results: Contact, Tan relations, sum rules}}}
\medskip

A remarkable consequence of the diluteness of cold atoms ($k_F r_0\ll 1$) is that we can take the zero-range 
($r_0\!\rightarrow\!0$) limit and obtain a number of \emph{exact} results that give nontrivial relations between various observable across the entire crossover.
These include the \emph{Tan relations}~\cite{tan08,Tan08en} and exact sum rules 
for a variety of spectral functions, which we discuss below.  An excellent review of these universal relations is given in Ch.~6 in Ref.~\cite{zwerger2011bcs}.    

The central quantity underlying these exact identities is the {\it contact} $C$~\cite{Tan08en,braaten08prl} that quantifies the probability of 
finding two particles of opposite spin close together. One way to introduce $C$ is via the short distance structure of the two-particle correlator~\cite{Tan08en,Zhang09}
\begin{equation}
\rho_{2}(r \ll k_F^{-1})
\equiv \int d^3\bR \langle {\psi}^{\dagger}_{\uparrow}(\bR\!+\!\tfrac{\br}{2})
{\psi}^{\dagger}_{\downarrow}(\bR\!-\!\tfrac{\br}{2})
{\psi}_{\downarrow}(\bR\!-\!\tfrac{\br}{2}){\psi}_{\uparrow}(\bR\!+\!\tfrac{\br}{2})\rangle
\approx \frac{C}{16\pi^2}\left({1}/{r}-{1}/{a}\right)^2.
\label{rho2}
\end{equation}
The form of this result is dictated by two-body physics, since at short distances $\rho_2$ is just the square of the two-particle 
wavefunction in Eq.~(\ref{2particle}). But the prefactor $C$ contains information about many-body physics.  
The contact $C$ also determines the large-$k$ ``tail" of the momentum distribution
of spin-$\sigma$ fermions
\beq n_{\sigma}(k\gg k_F) \to C/k^4.
\label{nlargek}\eeq 
This shows that the contact must be of the form $C = k_F^4 \, {\cal C}({T/E_F},{1/k_F a})$.
Although the $n_{\sigma}(k)$ tail and related results have been obtained in a variety of different 
ways~\cite{tan08,Tan08en,Zhang09,wern09closed,comb09}, the operator product expansion~\cite{braaten08prl,braaten08pra}
leads to the most elegant derivation. 

While it might seem that the short-distance or large-$k$ behavior of correlations should be of limited interest,
Tan showed that the contact $C$ was directly related to a wide range of thermodynamic quantities.  
The first of the Tan relations that makes this connection is 
\beq \left({\partial {\cal{E}}}/{\partial a^{-1}}\right)_{S} = - {C}/{4\pi m},\label{adiabatic}\eeq
the so-called adiabatic relation, where ${\cal{E}}$ is the energy density and $S$ is the entropy.  It follows from an 
application of the Hellmann-Feynman theorem 
$(\partial {\cal{E}}/\partial\lambda)_S = \langle\partial \hat{H}/\partial \lambda\rangle =  
\int d^3\br \rho_2(r)\partial V(r;\lambda)/\partial \lambda$ to a Hamiltonian with a short-range interaction 
$V(r;\lambda)$, with $\lambda$ a microscopic parameter (e.g.,  the magnetic field detuning from a Feshbach resonance) 
which tunes the scattering length $a(\lambda)$.  
The two-particle Schr\"odinger equation gives $ \partial\lambda/\partial a^{-1} = -4\pi/[ m\int d^3\br (1/a-1/r)^2\partial V/\partial\lambda]$~\cite{Zhang09,yu09correlations}.  Using Eq.~(\ref{rho2}) we immediately obtain Eq.~(\ref{adiabatic}).  

Another Tan relation expresses the total energy as functional of $n_{\sigma}(k)$:
\beq E = \sum_{\bk,\sigma}\frac{k^2}{2m}\left[n_{\sigma}(\bk)-\frac{C}{k^4}\right] + \frac{C}{4\pi m a}.
\label{TanE}
\eeq
Using Eq.~(\ref{rho2}) to evaluate the interaction energy $V\equiv \int d^3\br V(r,\lambda)\rho_2(r)$ and making use of the fact that Eq.~(\ref{2particle}) solves the two-particle Schr\"odinger equation, one finds $V =  C/4\pi ma - C\Lambda/2\pi^2m$ where the cutoff $\Lambda\sim 1/r_0$.
The divergence in the interaction energy is exactly cancelled by the one in the kinetic energy, which must exist in view of
the tail of Eq.~(\ref{nlargek}), thus leading to a total energy (\ref{TanE}) which is cutoff independent.
Another Tan identity is 
\beq P = {2\cal{E}}/{3} + {C}/{12\pi ma},\eeq 
the pressure relation, which follows straightforwardly from the definition $P = -(\partial E/\partial\Omega)_{S,N}$ where $\Omega$ is the volume and the adiabatic relation (\ref{adiabatic}).  
We note that the Tan relations as well as the consistency of $C$ determined from 
various measurements, such as the $n_{\sigma}(k)$-tail and the
the radio frequency (RF) spectroscopy tail (see below), have been verified experimentally~\cite{stew10contact}.

The key insight underlying all Tan identities is that the form of the short-distance properties of Fermi gases are determined by two-body physics, with the overall strength controlled by the many-body contact $C$. The same is true for short-time properties or the high-frequency tails ($E_F \ll \omega \ll 1/mr_0^2$)  of various spectral functions, 
which are all proportional to $C/\omega^{\gamma}$ with $\gamma$ determined by two-body physics.
Examples include the RF spectral function tail
$I(\omega) \sim C/\omega^{3/2}$~\cite{pier09rf,schn10shortrange} and the tail of the long-wavelength dynamic structure factor
$S({\bf q},\omega) \sim C q^4/\omega^{7/2}$~\cite{tayl10visc,son10}. The large-$\omega$ RF tail has been used as one
of the ways to measure the contact~\cite{stew10contact} (for another way to measure the contact, see Ref.~\cite{hoinka13}), and the
tails play an important role in deriving sum rules, a topic which we turn to next.

\medskip
\noindent
{\bf Sum rules:} Sum rules have played a central role in condensed matter physics. These exact results --
derived using Kubo formulae, causality, and commutation relations -- 
are useful for analyzing experimental data, constraining approximate calculations, and deriving rigorous results.
For lack of space, we omit a discussion of the ``clock shift'' sum rule on the first-moment of the 
RF spectral function, which is proportional to the interaction energy~\cite{baym07,punk07rf,zhan08rf}.

We focus here on the sum rules for the spectral functions~\cite{tayl10visc} of the shear viscosity $\eta(\omega)$
and the bulk viscosity $\zeta(\omega)$ in 3D. The $\eta$-sum rule~\cite{enss11viscosity,tayl12visc}
\beq
\int^{\infty}_0 {d\omega \over \pi} \left[\eta(\omega)- {C}/{15\pi \sqrt{m\omega}}\right] ={\cal{E}}/{3} -{C}/{12 \pi m a}\label{3Dshear}
\eeq
relates the integral of a spectral function over all frequencies to thermodynamics.
Note that a large-$\omega$ tail of the form described above is subtracted out to
get a convergent, \emph{universal} answer~\cite{tayl12visc}, independent of cutoff in the $\Lambda \sim 1/r_0 \to \infty$ limit.
This sum rule has been useful in calculations of (the d.c.~value of) the shear viscosity $\eta \equiv \eta(\omega=0)$;
see next Section. It acts as a numerical check on conserving approximations~\cite{enss11viscosity} and as a constraint imposed
on the analytic continuation of QMC data~\cite{wlaz12viscosity} from imaginary time to real frequency.

The $\zeta$-sum rule~\cite{tayl10visc,tayl12visc} (which applies to $\zeta_2$ below $T_c$) is given by
\beq
\int^{\infty}_0 {d\omega \over \pi} \zeta(\omega) = P  - {\cal{E}}/{9} - {\rho c^2_{\bar{s}}}/{2} = (72\pi ma^2)^{-1}\left({\partial C}/{\partial a^{-1}}\right)_{\bar{s}},
\label{3Dbulk}
\eeq
where $c_{\bar{s}} = (\partial P/\partial\rho)^{1/2}_{\bar{s}}$ is the adiabatic sound speed with $\bar{s}= S/N$ the entropy per particle.
It follows from the $T/E_F$ scaling of thermodynamic functions that the right-hand side of (\ref{3Dbulk}) vanishes at unitarity.
The second law of thermodynamics implies $\zeta(\omega) \geq 0$. It then follows that 
$\zeta(\omega;T) \equiv 0$ at unitarity~\cite{tayl10visc}. This generalizes the result that $\zeta = 0$ in a 
scale-invariant system~\cite{son07bulk,wern06unitary} to all frequencies ($\omega \ll 1/mr_0^2$),
despite the presence of three energy scales: $E_F, T$ and the external $\omega$. Another interesting prediction that follows from
the vanishing of the bulk viscosity spectral function at unitarity is that~\cite{tayl10visc} 
$\eta(\omega) = \lim_{q\to 0} 3\omega^3 {\rm Im}\chi_{\rho,\rho}({\bf q},\omega)/4q^4$.
This should permit, in principle, a measurement of the shear viscosity spectral function
at unitarity using resonant Bragg scattering that measures dynamical density-density correlations (see Ref.~\cite{veeravalli08}).  The analogue of (\ref{3Dbulk}) in 2D has also been used~\cite{tayl12visc} to account for a small empirical bulk viscosity through the 2D BCS-BEC crossover~\cite{vogt12scaleinv}.  

\medskip
\noindent
{\underline{{\bf 6. Thermodynamics, transport and spectroscopy at unitarity}}}
\medskip

We have already noted the pioneering cold atoms experiments on the BCS-BEC crossover in the Introduction.
We now describe three beautiful new experiments, and related theory, that shed light on different aspects of the unitary Fermi gas: 
(1) Thermodynamic measurements~\cite{ku12} at unitarity that clearly show a signature of the
superfluid phase transition in the specific heat and also permit detailed comparison with high precision QMC~\cite{vanhoucke12}.
(2) Viscosity measurements~\cite{cao11viscosity} at unitarity, which find evidence for a very strongly interacting system with a shear viscosity to entropy density ratio $\eta/s$ that comes close to saturating a conjectured bound.
(3) RF spectroscopy experiments~\cite{stew08arpes,gaeb10pseudo}, analogous to angle-resolved photoemission (ARPES), suggestive of a pseudogap 
arising from pairing in the normal state.
We will not discuss here other important experiments and related theories, e.g., collective oscillations of 
trapped gases~\cite{stri04,Tey13,sido132ndsound}, including the remarkable recent observation~\cite{sido132ndsound} of second sound~\cite{taylor09tfl} and measurement of the superfluid density.

\begin{figure}   
\begin{center}
\includegraphics[width=0.4\textwidth]{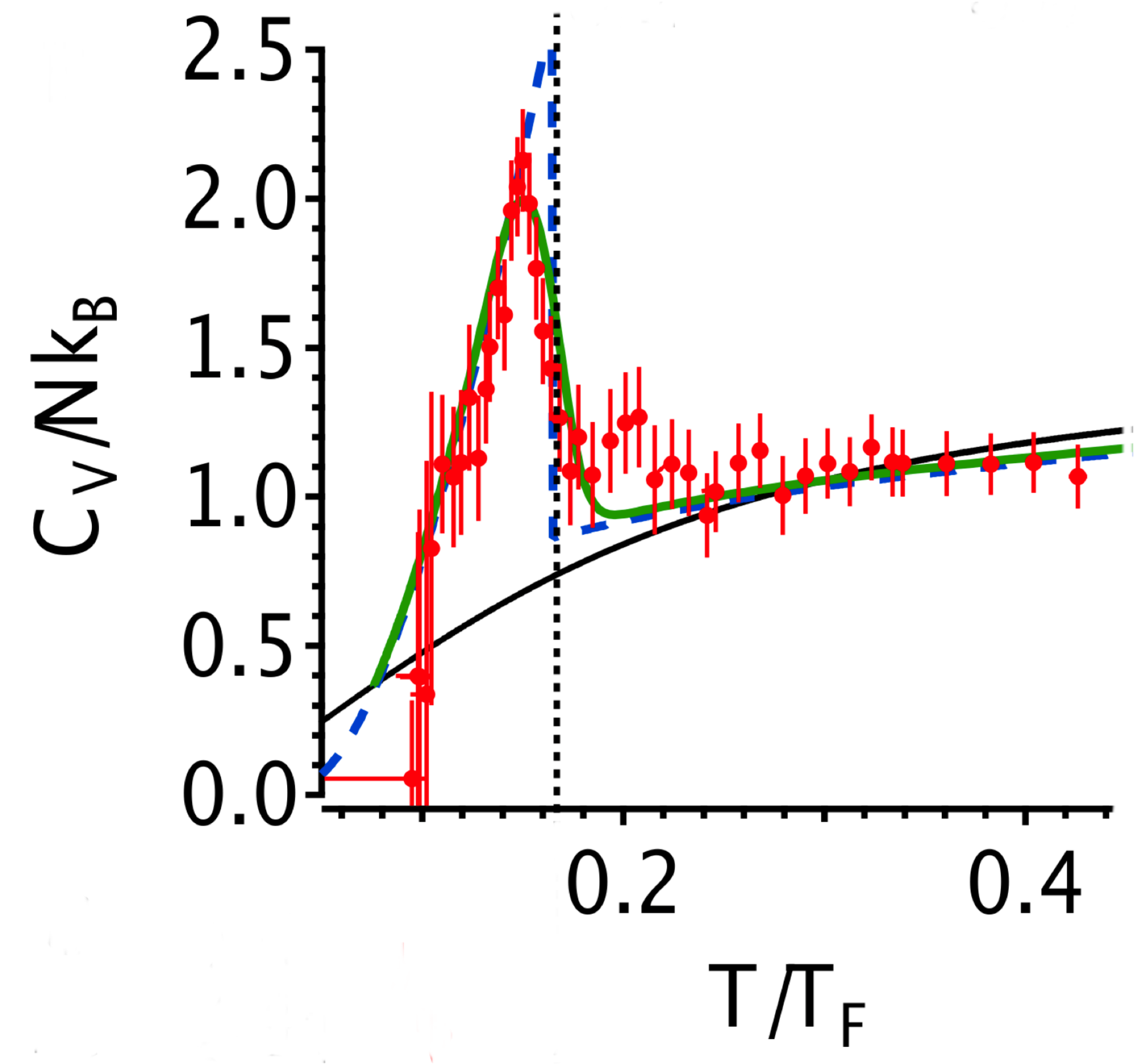}
\caption{The specific heat of the unitary Fermi gas as a function of $T/F_F$ exhibits a phase transition
to the superfluid state at $T_c\simeq 0.18E_F$. The red dots denote experimental points; the colored curves are various theoretical results for comparison.  From Ref.~\cite{ku12}.  
} \label{thermofig}
\end{center}
\end{figure}

\medskip
\noindent
{\bf Thermodynamics of the unitary Fermi gas:} The harmonic trap $V_{\mathrm{trap}}(\br) = {m\over 2}\sum_i \omega^2_i\br_i^2$, $(i=x,y,z)$ 
leads to an inhomogeneous density $n(\br)$. The connection between the properties of a trapped gas and that of a uniform system can 
be established by using the Thomas-Fermi local density approximation (LDA)~\cite{gior08review}, where
one uses a ``local chemical potential'' 
\beq \mu(\br) \equiv \mu[n(\br)] = \mu_0 - V_{\mathrm{trap}}(\br).
\label{TFA}\eeq
Here $\mu_0$ is the thermodynamic chemical potential of the trapped gas, fixed by the condition $N=\int d\br n(\br)$, and $\mu[n] \equiv (\partial {\cal E}/\partial n)_{S,N}$ is the chemical potential of the uniform gas, determined by the energy density ${\cal E}$ of the uniform system.  A number of experiments~\cite{kina05heat,stew06pot,luo07entropy,nasc10thermo,horikoshi10thermo} 
have measured the energy of the unitary Fermi gas in a trap, given within LDA by
$E_{\mathrm{trap}}= \int d^3\br\, {\cal{E}}\left[T/E_F(\br)\right]$.  $T/E_F(\br)$ varies from its lowest value at the trap centre to infinity 
at the edge of the cloud [$E_F(\br) \sim n^{2/3}(\br)$] where the density vanishes.
Thus  $E_{\mathrm{trap}}(T)$ is effectively a weighted sum over all temperatures, greatly complicating
the task of extracting the equation of state ${\cal{E}}(T)$ of the uniform gas from it.

A crucial step in obtaining \emph{local} thermodynamic quantities was 
the realization~\cite{zhou09,ho09thermo} that the measured density profile $n(\br)$ yields the local isothermal compressibility $\kappa_T(\br)$.
Using Eq.~(\ref{TFA}),  $\kappa_T(\br) =  n^{-2}(\br)(\partial \mu[n(\br)]/\partial n)^{-1}_T$ can be related to spatial derivatives
of $n(\br)$ and  $V_{\mathrm{trap}}(\br)$.
Remarkably, using only standard thermodynamic relations and scale invariance at unitarity~\cite{ho04uni}, 
(local) thermodynamic quantities, such as the pressure, energy, chemical potential, entropy and specific heat, 
can all be determined from $\kappa_T$~\cite{ho09thermo,vanhoucke12,ku12}.  As an example
we show in Fig.~\ref{thermofig} the specific heat~\cite{ku12} of a unitary Fermi gas, which gives clear evidence for the superfluid phase transition, known to be in the 3D XY universality class. 

The same experiments also led to the most precise measurement of $\xi_s=E_0/({3\over 5}NE_F)$ (see Table I) and a
detailed comparison of normal state thermodynamics with new numerical techniques like ``bold diagrammatic" QMC~\cite{vanhoucke12}.
The remarkable agreement between experiment and QMC without adjustable parameters, shows the potential for quantum gases to benchmark novel computational techniques.

\medskip
\noindent
{\bf Viscosity:}  Transport in strongly interacting quantum systems without well-defined quasiparticles
is a subject of central interest in diverse areas of physics. In quantum materials the focus has been on electrical transport in non-Fermi liquid
regimes, often near quantum critical points. Here we look at the viscosity of the unitary Fermi gas, which -- surprisingly -- has  
connections with the quark-gluon plasma (QGP) and string theory!  

Much of the interest in the shear viscosity $\eta$ of the unitary Fermi gas and the QGP produced in heavy ion collisions 
originates in a gauge-gravity duality calculation~\cite{poli01viscosity} of $\eta/s=\hbar/4\pi k_B$, where $s$ is the entropy density (with $\hbar$ and $k_B$ restored for clarity),
subsequently conjectured to be a universal lower bound~\cite{kovt05visc}.  Remarkably, there are no known experimental violations, which raises the questions of whether quantum mechanics places bounds on quasiparticle lifetimes and transport. The two fluids that come closest to saturating the bound are the
unitary Fermi gas and the QGP~\cite{scha09perfect}, despite the many orders-of-magnitude difference in their temperature and density. 
The property that these two systems share is that both are very strongly interacting, and both exist in a regime where standard Boltzmann transport
theory is of questionable validity. In kinetic theory, $\eta\sim n p \ell$, where $n$ is the density of quasiparticles, $p$ their characteristic momentum
and $\ell$ the mean free path. Using $s\sim nk_B$, we see that any fluid that comes close to the conjectured lower bound will necessarily
violate $p\ell \gg \hbar$, the condition for well defined quasiparticles.

 \begin{figure}   
\begin{center}
\includegraphics[width=0.9\textwidth]{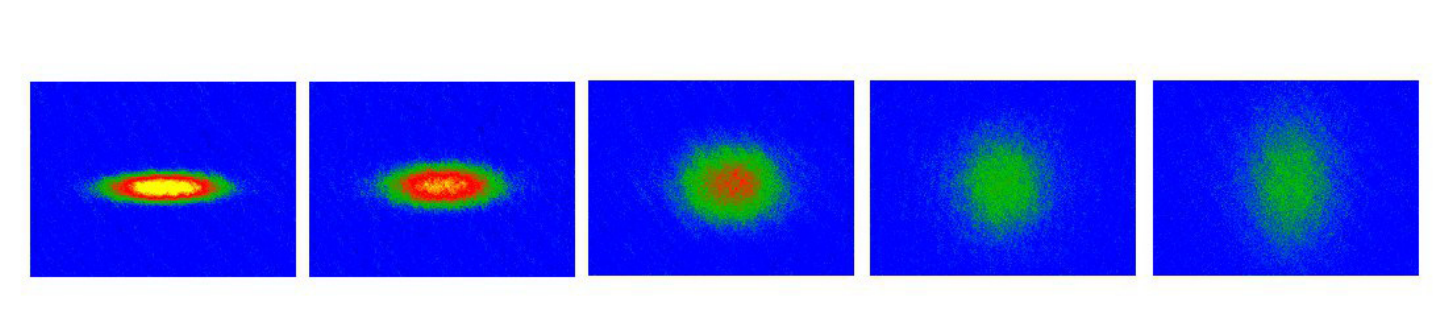}
\caption{Elliptic flow of a strongly-interacting Fermi gas close to unitarity.  The tight radial confinement produces a large pressure gradient in this direction, yielding flow predominantly along this axis for a low viscosity system.  From Ref.~\cite{cao11viscosity}.  
}
\label{ellipticfig}
\end{center}
\end{figure}

The first indication that the unitary Fermi gas might have very low viscosity came from the
observation of ``elliptic flow"~\cite{ohar02science} in its expansion dynamics once the anisotropic
(cigar shaped) trap is turned off; see Fig.~\ref{ellipticfig}. This is very similar to the behavior seen 
in heavy ion experiments. Elliptic flow is strictly a property of an \emph{ideal} fluid, with zero viscosity, wherein flow follows pressure gradients,
in accordance with Euler's equation $\rho(\partial_t + \bv\cdot\nabla)\bv = -\nabla P$.  
For the unitary gas,  the non-zero shear viscosity $\eta$ (recall that the bulk viscosity vanishes)
leads to corrections that can be modeled using hydrodynamics.
That ultracold gases---some six orders of magnitude more dilute than air---should obey hydrodynamics
over the typical dynamical timescale in experiments, set by the inverse trap frequency $\omega^{-1}_{\mathrm{trap}} \sim 10$ms, might seem counterintuitive. However, strong interactions near unitarity imply a very short time scale
$\tau_R\sim 1/E_F$~\cite{massignan05visc,bruu07viscosity} to reach local equilibrium.
(Far from unitarity, $\tau_R$ is large and one enters a collisionless regime,
leading to ballistic expansion instead of elliptic flow~\cite{ohar02science,gior08review}).

Recent experiments~\cite{turl08perfect, cao11viscosity} have given quantitative estimates of the shear viscosity in a unitary Fermi gas 
by detailed modeling of the elliptic flow and by measuring the damping of collective oscillations. They find a minimum value of 
$\eta k_B/\hbar s\simeq 0.2$ just above $T_c$. Calculations on the unitary Fermi gas also find that $\eta$ is minimum 
above $T_c$ with conserving approximation estimates $\eta/n\hbar\!\simeq\!0.5;\, \eta k_B/\hbar s\!\simeq\!0.6$~\cite{enss11viscosity} 
and QMC values $\eta/n\hbar\!\simeq\!0.18;\, \eta k_B/\hbar s\!\simeq\!0.2$~\cite{wlaz12viscosity}. All calculations need
numerical schemes for analytic continuation of imaginary time results, and there is ongoing
discussion about their accuracy~\cite{chafin12viscmin,romatschke12viscmin}.
By way of comparison, first generation experimental values for the QGP in the heavy ion collisions
are around $ \eta k_B/\hbar s\!\simeq\!0.5$~\cite{scha09perfect}. 

\medskip
\noindent
{\bf RF spectroscopy:} Measuring the pairing gap across the BCS-BEC crossover has proved to be harder than
might have been expected in a field that specializes in high-precision spectroscopy. The basic idea is to absorb a
radio frequency (RF) photon and make a transition from one of the states $|\sigma\rangle$ ($\sigma = \uparrow, \downarrow$) 
involved in pairing to a third hyperfine state $|f\rangle$. Early experiments were difficult to interpret because of strong interactions between the final state $|f\rangle$ and 
atoms in $|\sigma\rangle$ (in $^6$Li, but not in $^{40}$K) as well as trap averaging over an inhomogeneous density.
Even after these problems are ameliorated by appropriate choice of hyperfine states and use of local imaging,
the RF absorption threshold~\cite{torma00,kett08varenna} measures $E_{\rm th} = \sqrt{\Delta^2 + \mu^2} - \mu$ (within MFT), and not the gap $\Delta$.
RF photons excite atoms in all ${\bk}$ states and $E_{\rm th}$ is determined by ${\bk} = 0$ fermions, instead of ones close to $k_F$.
To measure $\Delta$, unpaired atoms were injected into the superfluid to create a slightly spin-imbalanced mixture~\cite{schi08gap}. 
The response of the paired and unpaired atoms at different frequencies then led to the estimate of the 
the pairing gap $\Delta=0.44 E_F$ at unitarity (see Table I).

An exciting new development is {\it momentum resolved} RF spectroscopy~\cite{stew08arpes,gaeb10pseudo}, 
analogous to angle-resolved photoemission spectroscopy (ARPES)~\cite{dama03arpes, camp04arpes},
a powerful probe of quantum materials. This gives direct information about the single-particle spectral function
$A_{\sigma}(\bk, \omega)\!=\!-\mathrm{Im}G_{\sigma}(\bk,\omega+i0^{+})/\pi$, a quantity of central interest in 
many-body physics. A measurement of the final state momentum distribution, together with
kinematic constraints, leads to 
$I({\bk}, \omega) \propto n_F(\epsilon_{\bk}-\mu_{\sigma}-\omega) A_{\sigma}(\bk,\epsilon_{\bk}-\mu_{\sigma}-\omega)$.
Here $\omega \equiv \omega_{\mathrm{RF}} - \Delta\omega$, with $\omega_{\mathrm{RF}}$ the frequency of the applied RF field and 
$\Delta\omega$ the frequency difference between the $|\sigma\rangle$  and $|f\rangle$ hyperfine states,
$\epsilon_{\bk} = k^2/2m$ and the Fermi function $n_F$  enters since only occupied $|\sigma\rangle$ states contribute to the signal. 
(The usual RF spectroscopy  measures $I(\omega)\equiv \sum_\bk I({\bk}, \omega)$).
 
\begin{figure}   
 \begin{center}
\includegraphics[width=0.9\textwidth]{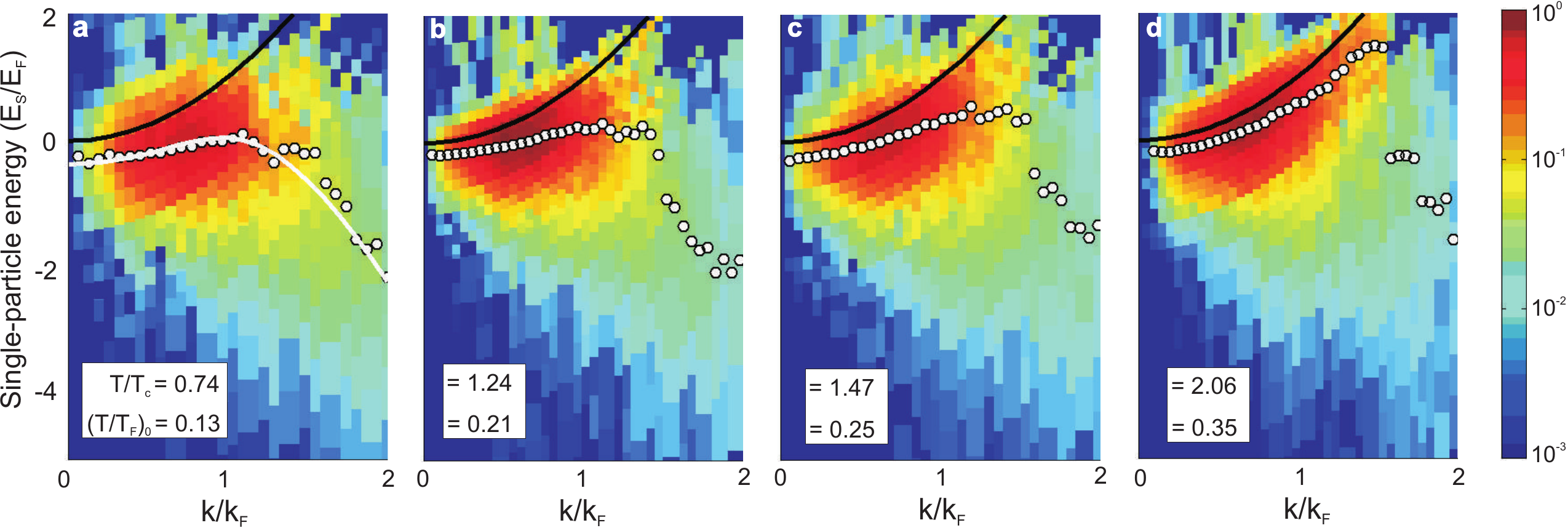}
\caption{Momentum resolved RF spectroscopy of the unitary Fermi gas for a range of temperatures through the superfluid (left) to normal (right) transition.  White dots denote points of highest spectral intensity; the black curve is the quadratic dispersion for a free Fermi gas, and the white curve if a fit to a BCS-like dispersion.
From Ref.~\cite{gaeb10pseudo}
}
\label{RFfig}
\end{center}
\end{figure}

The $\bk$-resolved RF response~\cite{gaeb10pseudo} at unitarity in Fig.~\ref{RFfig} clearly shows a 
``back-bending" (particle-hole mixing) of the dispersion characteristic of a gapped Bogoliubov quasiparticle below $T_c$.
A precise estimate of the gap is not available at this time, as it requires knowledge of the chemical potential.
A dramatic aspect of this data is the persistence of back-bending above $T_c$, which
is tempting to associate with a pairing pseudogap~\cite{rand98varenna}. 
However, as shown in Ref.~\cite{schn10shortrange}, there is a universal large-$k,|\omega|$ tail
in $A_{\sigma}(\bk, \omega)$ (related to the large-$\omega$ tail in $I(\omega)$ mentioned in the previous Section) that has its origin in the short-range physics of dilute gases and which \emph{guarantees} the existence of incoherent spectral weight that exhibits back-bending, independent 
of whether the system has a gap or not. Thus it becomes important to use back-bending near $k_F$ as a signature
of a pseudogap for $T_c\!<\!T\!<T^*$, as distinct from back-bending at large $k\gg k_F$ that would exist even above $T^*$
in the absence of a pseudogap. The data in Fig.~\ref{RFfig} are consistent with this expectation insofar as  the near-$k_F$ behavior evidences a conventional single-particle dispersion at the highest temperatures.

\medskip
\noindent
{\underline{{\bf 7. Related problems and open questions:}}}
\medskip

\noindent
{\bf Pairing pseudogap:} As discussed above, the normal 
state evolves smoothly from a Fermi liquid at weak coupling to a normal Bose gas
at strong coupling, and does so by exhibiting a pseudogap~\cite{rand92,triv95,rand98varenna} 
in the range $T_c\!<\!T\!<\!T^*$, where $T^*$ is the the pairing crossover scale; see Fig.~\ref{f:phasediagram}.
We use the word ``pseudogap'' to describe the effects of incoherent pairing above $T_c$
that lead to strong suppression of low-energy spectral weight for single-particle excitations.
Its observable consequences include a gap-like dispersion and associated anomalies in the
density-of-states and spin susceptibility~\cite{rand92,triv95,rand98varenna} that show an
unusual $T$-dependent suppression.  We discuss here the theoretical and experimental 
evidence for such anomalies in the unitary regime,
and then turn to the extent to which these considerations relate to the more complex set of
phenomena observed in high $T_c$ superconductivity in cuprates.

A crucial question is the temperature range over which pseudogap anomalies
are observed at unitarity. Different pair-fluctuation approximation schemes give
differing answers~\cite{haus09rf,chie20pg}.  The best QMC estimate of the pairing temperature 
$T^* \simeq 0.2 E_F$~\cite{magi09gap,bulg11pg,wlaz13cp} at unitarity 
is considerably lower than the MFT $T^*$ (Fig.~\ref{f:phasediagram}), but nevertheless
larger than $T_c \simeq 0.15 E_F$. Moreover, as one moves to the BEC side of unitarity, 
the pseudogap regime grows (although, at sufficiently strong coupling, the normal
state ceases to be quantum degenerate, and the pseudogap is simply the molecular binding energy).
We note that the analysis of spectral properties in QMC and in diagrammatic approximations 
is complicated by the challenging problem of analytic continuation of numerical data from imaginary to real frequencies.

The experimental situation is not entirely clear. The angle-resolved RF data~\cite{gaeb10pseudo}
are highly suggestive of a pseudogap, despite the caveats discussed above. 
Another aspect of the deviations from Fermi liquid behavior in the pseudogap regime 
is that the sharp Bogoliubov quasiparticle peaks (below $T_c$) should be greatly broadened above $T_c$. Although the present RF energy
resolution is not sufficient to address this question, the extremely small
shear viscosity~\cite{cao11viscosity} is consistent with a short lifetime for excitations just above $T_c$.
On the other hand, spin susceptibility $\chi_s$ measurements~\cite{somm11,nasc11LFL} do not give clear evidence for characteristic $T$-dependent suppression, with $d\chi_s/dT>0$~\cite{rand92,triv95,rand98varenna,wlaz13cp}.
The question of a pseudogap in 2D, where pairing is stronger and $T_c$ is suppressed, 
also remains to be investigated.


Early work on the pairing pseudogap~\cite{rand92,triv95,rand98varenna} 
was motivated by the question: does the normal state of a short coherence length
superconductor, with pair size comparable to
interparticle distance, show deviations from Fermi liquid behavior?
It is thus fitting to conclude this discussion by comparing the pseudogap 
in the BCS-BEC crossover with the (still not well understood)
pseudogap in the high-$T_c$ cuprate superconductors~\cite{lee06hightc,dama03arpes,camp04arpes}.

The cuprates differ from the BCS-BEC crossover discussed here
in their dominant interactions (Coulomb repulsion {\it vs.}~s-wave attraction), pairing symmetry (d-wave {\it vs.}~s-wave), 
and effective dimensionality (quasi-2D layers {\it vs.}~3D).
The highly anisotropic pseudogap in underdoped cuprates
is likely impacted by proximity to the antiferromagnetic Mott insulator,
by incoherent pairing above $T_c$, and by competing orders such as charge density waves. The strongly interacting Fermi gas is a simpler problem 
with a single instability to $s$-wave pairing. 
Thus if a gap exists above $T_c$, it can only be related to precursor pairing correlations.

\medskip
\noindent
{\bf Spin-imbalanced Fermi gases:}  In this review we have only considered Fermi gases with
equal densities of the two spin species.  
A very fertile area of theoretical and experimental activity is the {\it imbalanced} case with
$n_{\uparrow}\neq n_{\downarrow}$~\cite{shee06phase,pari07finitetemp,bulg08lo,chevyreview11,gubb12}.   
Spin imbalance acts as a Zeeman field, and may open the door to the possible realization 
of the spatially modulated superfluid states proposed by
Fulde--Ferrell--Larkin--Ovchinnikov (FFLO)~\cite{FF64,LO64}.  Such states are of considerable interest in both condensed matter physics~\cite{Won04,Mart05} and in color superconductivity in quantum chromodynamics~\cite{alfo00color}.  
Experiments have made great progress in mapping out the temperature-imbalance phase diagram 
at unitarity~\cite{shin07phasediagram}, but have not seen the FFLO state in 3D. Experiments have found signatures of FFLO in
1D~\cite{liao10Fermi1D}, where it is predicted to exist over a large range of parameters~\cite{orso07fflo}. 
There has also been much progress on the problem with large spin-imbalance, which leads to a Landau Fermi liquid of ``polarons''~\cite{lobo06polaron,chev06polaron,prok08polaron}
whose properties have also been measured~\cite{schi09polaron,nasc10thermo,nasc11LFL}.

\medskip
\noindent
{\textbf{Two dimensions:}} 
The 2D BCS-BEC crossover~\cite{rand89bound,rand90,petr03quasi2d,bert11} is of great interest
since 2D is the marginal dimension both for the formation of quantum bound states and for
classical fluctuations of the superfluid order parameter.  Recent experiments~\cite{somm12,frol12,vogt12scaleinv} have made a number of
intriguing observations.
The RF absorption threshold in 2D~\cite{somm12} is found to be just the dimer binding energy $E_b$, a mean field theory
(MFT) prediction~\cite{rand89bound} that one would not have expected to be quantitatively valid in 2D~\cite{randeria12}.
An undamped, monopole breathing mode is found~\cite{vogt12scaleinv} to oscillate at twice the trap frequency for
a  broad range of temperatures and couplings across the 2D crossover. This apparent scale-invariant behavior,
in a theory with an explicit scale $E_b$, is also very surprising~\cite{tayl12visc}, but this too emerges naturally from the
$T\!=\!0$ MFT. An important open question then is to understand why the effects of quantum and thermal fluctuations 
are so weak for some observables, though not all, in 2D.

\medskip
\noindent
{\textbf{Non \textit{s}-wave superfluids:}} In marked contrast to the smooth $s$-wave crossover, the higher angular momentum pairing
problem~\cite{legg80,rand90,ho05pwave,ohashi05pwave} can have a topological quantum phase transition separating the weak and strong coupling phases~\cite{read00,klin04,gura05pwave}. At this time, there appear to be a number of technical challenges to realizing 
the $p$-wave BCS-BEC crossover experimentally~\cite{gaeb07pwave}. However, it would be very exciting to realize the
chiral $p_x+ip_y$ superfluid state in 2D~\cite{gura05pwave}, which
supports Majorana excitations~\cite{read00,klin04} that are important for topologically protected quantum computation.
 

\medskip
\noindent
{\underline{{\bf 8. Concluding thoughts:}}}
\medskip

A most exciting and unique aspect of ultracold atomic gases is the ability
to tune parameters in simple Hamiltonians with exquisite precision, over timescales which are essentially instantaneous compared to the natural dynamical timescales in the problem, $\omega^{-1}_{\mathrm{trap}}\sim  10$ ms, $E^{-1}_F\sim 100$ $\mu$s,
and access very strongly interacting regimes.
This is in sharp contrast to quantum materials, where the possibility of
tuning interactions is more limited and strong-correlations come 
hand in hand with other complexities like phonons and disorder.
While this leads to an infinitely rich phenomenology in the solid-state,  it is
often resistant to simple quantitative modeling.
Thus ultracold gases presents us with an unprecedented opportunity to explore
effects that arise purely from strong interactions, unencumbered by other complications. 

The Fermi gas with interaction tuned via a Feshbach resonance 
realizes the BCS-BEC crossover, a problem of long standing interest. 
In its strongly interacting unitary regime, it gives new insights into the problems of high-$T_c$ superfluidity, the pairing pseudogap, and transport
without sharp quasiparticles. The fact that one can see superfluidity in a Fermi system
with a $T_c$ that is $15-20 \%$ of the bare Fermi energy  $E_F$ is itself remarkable and raises the following question: Is there an upper bound on $T_c/E_F$?

The study of the unitary Fermi gas has brought together researchers from diverse areas of physics.  As a result of this cross-fertilization, new theoretical tools have been brought to bear on the unitary Fermi gas, 
 which in turn exhibits remarkable universal properties that shed light on other strongly interacting problems.
These include nuclear physics and quantum chromodynamics on the one hand, and
 gauge-gravity duality in string theory on the other, which have interest in
 scale-invariant (or nearly scale-invariant) systems exhibiting an unusually low
 shear viscosity. Although its gravity dual is not known, it is an interesting open
 question whether holographic techniques will give new insight into non-relativistic
 systems such as the unitary Fermi gas.

 \bigskip
\noindent
{\bf Acknowledgements:} We would like to thank Eric Braaten, Roberto Diener, Allan Griffin, Jason Ho, Randy Hulet, Deborah Jin,
Wolfgang Ketterle, Tony Leggett, Lev Pitaevskii, Rajdeep Sensarma, Sandro Stringari, Joseph Thywissen, Nandini Trivedi, Shizhong Zhang,  Willi Zwerger and Martin Zwierlein
for numerous discussions.
MR gratefully acknowledges the National Science Foundation and the
Army Research Office for funding his research on ultracold Fermi gases; he is currently supported by NSF-DMR-1006532.  ET acknowledges funding from NSERC and the Canadian Institute for Advanced Research (CIFAR).

 \newpage

\bibliographystyle{unsrt}
\bibliography{References}

\end{document}